\documentclass[runningheads]{llncs}
\usepackage[T1]{fontenc}
\usepackage{graphicx}
\usepackage{xcolor}

\usepackage{booktabs}

\usepackage{acronym}
\usepackage{microtype}              
\usepackage{subcaption}
\usepackage{cleveref}

\usepackage{multirow}

\newacro{ai}[AI]{Artificial Intelligence}
\newacro{hpc}[HPC]{high-performance computing}
\newacro{llm}[LLM]{Large Language Model}
\newacro{cv}[CV]{Computer Vision}
\newacro{resnet}[ResNet]{Residual Networks}
\newacro{vit}[ViT]{Vision Transformer}
\newacro{ldm}[LDM]{Latent Diffusion Model}
\newacro{tdp}[TDP]{Thermal Design Power}

\begin{document}
\title{AI Application Benchmarking: Power-Aware Performance Analysis for Vision and Language Models}
\titlerunning{Power-aware AI Benchmarking}
\author{Martin Mayr\inst{1}\orcidID{0000-0002-3706-285X} \and
Sebastian Wind\inst{1}\orcidID{0009-0004-7859-5785} \and
Lukas Schröder\inst{1,2}\orcidID{0009-0001-2842-568X} \and
Mohammadmoein Moradi\inst{1}\orcidID{0009-0004-6643-5186} \and
Georg Hager\inst{1}\orcidID{0000-0002-8723-2781}\and
Harald Köstler\inst{1,2}\orcidID{0000-0002-6992-2690}\and
Gerhard Wellein\inst{1}\orcidID{0000-0001-7371-3026}
}
\authorrunning{M. Mayr et al.}
\institute{Erlangen National High Performance Computing Center (NHR@FAU), Friedrich-Alexander-Universität Erlangen-Nürnberg, 91058 Erlangen, Germany \and
Chair of Computer Science 10, Friedrich-Alexander-Universität Erlangen-Nürnberg, 91058 Erlangen, Germany}

\maketitle              %
\begin{abstract}
Artificial Intelligence (AI) workloads drive a rapid expansion of high-performance computing (HPC) infrastructures and increase their power and energy demands towards a critical level. AI benchmarks representing state-of-the-art workloads and their understanding in the context of performance-energy trade-offs are critical to deploy efficient infrastructures and can guide energy efficiency measures, such as power limiting.  
We introduce a benchmarking framework with popular deep learning applications from computer vision (image classification and generation) and large language models (continued pre-training and inference) implementing modern methods. Our performance analysis focuses on throughput rather than ``time to completion'', which is the standard metric in HPC. We analyse performance and energy efficiency under various power-limit settings on NVIDIA H100, NVIDIA H200, and AMD MI300X GPUs. 
Our results reveal that no universal optimal power limit exists, as the efficiency peak varies across application types and GPU architectures. Interestingly, the two NVIDIA GPUs which mainly differ in their high-bandwidth memory (HBM) configuration show qualitatively different performance-energy trade-offs.
Code is available on Zenodo (\url{https://zenodo.org/records/20083679}) and GitHub (\url{https://github.com/RRZE-HPC/hpc-ai-perf-bench}).
 
\end{abstract}
\section{Introduction}
\label{sec:introduction}

Modern \ac{ai} research has undergone a structural shift toward increasingly large and complex models, characterized by growing model size, memory footprints, and dataset scale. 
Moreover, the key procedures of modern AI workflows, such as large-scale hyperparameter optimization, architecture search, and ablation studies, require massive throughput within feasible time frames.
These developments put small laboratories and single-node systems at a significant competitive disadvantage, as state-of-the-art AI training and evaluation increasingly exceed their available hardware compute and memory resources to train large models and to even fit them into GPU VRAM.
As a result, \ac{hpc} infrastructures become the primary providers of the centralized hardware resources to enable large-scale \ac{ai} research.

Despite this growing dependency, existing benchmarking approaches are poorly aligned with the practical performance requirements of deep learning workloads on HPC systems. 
Many established benchmarks, like MLPerf~\cite{mattson2019mlperf}, focus on time-to-convergence metrics
rather than throughput-oriented measurements (e.g., images or tokens per second). 
Even under fixed random seeds, convergence-based metrics can lead to non-deterministic and non-comparable results due to stochastic training dynamics. 
Moreover, scaling to multiple nodes introduces changes in effective batch size through data-parallel aggregation, often requiring specialized optimizers (e.g. LARS~\cite{you2017large} or LAMB~\cite{you2020large}) that inject additional noise into the optimization process. 
Existing large-scale benchmarking frameworks, such as MLPerf, further introduce significant system complexity and integration overhead, particularly in containerized environments without substantial system-specific adaptations.
The focus on performance metrics also frequently ignores the rapidly growing energy footprint of \ac{ai} workloads, which has become a critical concern at a global scale~\cite{chen2025datacenter}, with some hyperscalers even deploying gas turbines to meet electricity demand\footnote{\url{https://www.theguardian.com/environment/2026/feb/13/elon-musk-xai-datacenters-air-pollution-mississippi}} or consider to deploy small modular reactors\footnote{\url{https://world-nuclear.org/news-and-media/press-statements/world-nuclear-association-welcomes-microsoft-corporation-as-newest-member}}. 
This raises the need to better understand the trade-offs between raw performance and power consumption, as marginal performance gains often come at disproportionate energy costs, making the identification of power-efficient operating regimes critical for sustainable \ac{ai} deployment.

In this work, we introduce a work-in-progress \ac{ai} application benchmarking framework with first application scenarios and analyse their performance characteristics in combination with their energy profile. Specifically, we provide these contributions:
\begin{enumerate}
    \item \textbf{Open-source benchmarking framework} for performance evaluation of widely used AI-workloads in computer vision and large language models.
    \item \textbf{Systematic performance analysis} of the AI-workloads under different power-limit settings.
     \item \textbf{Energy efficiency analysis} in the context of performance--energy trade-offs.
    \item \textbf{Cross-platform experimental evaluation} across state-of-the-art GPU generations, including NVIDIA H100, NVIDIA H200, and AMD MI300X.
\end{enumerate}

\section{Related Work}
\label{sec:related_work}

The increasing energy demand of deep learning has motivated a broad body of work on efficient and sustainable AI. 
Survey papers summarize this landscape from complementary perspectives, covering compact models, efficient training and inference strategies, hardware-aware optimization, and broader lifecycle considerations for reducing energy use and emissions \cite{mehlin2023energy,menghani2023efficient,xu2021survey}.
Beyond surveys, several empirical studies quantify how model design and execution choices affect energy consumption. 
Xu et al. studied convolutional neural network training and showed that architectural choices should be evaluated not only by predictive quality but also by their energy use and emissions \cite{xu2023energy}. 
Tripp et al. introduced the BUTTER-E dataset, a large empirical benchmark with wattmeter-based energy measurements, and used it to analyze how network structure, model size, and hardware characteristics interact in non-trivial ways \cite{tripp2024measuring}. 
Most closely related to our motivation, Huber et al. investigated parallel neural network training and showed that energy consumption scales roughly with GPU resources, while the scaling still depends strongly on model, hardware, and parallelization configuration \cite{huber2025energy}. 
Together, these works establish energy efficiency as an important objective alongside performance and model quality.

Another line of related work studies runtime power management, in particular power capping, as a practical way to improve energy efficiency under system-level power constraints. 
Patki et al. compared GPU power capping and frequency capping in a scientific workflow and found that power capping can improve power efficiency while preserving performance more reliably \cite{patki2019comparing}. 
Allen et al. studied performance optimization in power-bounded GPU computing and proposed heuristics for selecting favorable operating points under a power budget \cite{allen2020performance}. 
More recent work extends this perspective to exascale systems: Patrou et al. evaluated power-capping metrics for an exascale science application on integrated CPU and GPU hardware, while Costa et al. characterized the impact of GPU power-management policies across multiple accelerator benchmarks and highlighted workload-dependent trade-offs between performance and energy savings \cite{patrou2025power,costa2025characterizing}.

However, existing literature focuses primarily on either general deep learning energy analysis or power-management studies for traditional HPC applications. In contrast, we study throughput-oriented benchmarking of representative modern AI workloads, including computer vision, image generation, large language model training, and inference, across multiple GPU platforms and power-limit settings.

\section{AI Application Benchmark}
\label{sec:methodology}

\begin{table*}[t]
\centering
\small
\caption{\ac{ai} workloads included in the proposed benchmarking framework.}
\label{tab:methodology_workloads}
\setlength{\tabcolsep}{7pt}
\begin{tabular}{p{1.6cm}p{2.3cm}p{4.4cm}p{1.6cm}}
\toprule
Domain & Task & Model & Mode \\
\midrule
CV & Classification & ResNet-50 \cite{he2016deep} & Training \\
CV & Classification & ViT-L/16 \cite{dosovitskiy2021image} & Training \\
CV & Generation & Stable Diffusion v2 \cite{rombach2022latent} & Training \\
LLM & Pre-training & LLaMA~3 8B, LitGPT \cite{touvron2023llama,grattafiori2024llama} & Training \\
LLM & Inference & LLaMA~3 8B, SGLang \cite{zheng2024sglang} & Inference \\
\bottomrule
\end{tabular}
\setlength{\tabcolsep}{6pt}
\end{table*}

We benchmark representative \ac{ai} tasks from \ac{cv} and \ac{llm}, as summarized in \Cref{tab:methodology_workloads}.
The tasks were selected to reflect common user behavior and frequently used workloads on our \ac{hpc} \ac{ai} clusters.
To focus on hardware-limited behavior, we report throughput rather than time-to-convergence, using images per second for \ac{cv} and tokens per second for \ac{llm}.
The modular framework allows models and implementations to be exchanged with minimal effort.
This workload mix spans discriminative and generative settings and enables comparison of compute-intensive and memory-sensitive execution patterns across accelerator platforms.

\paragraph{Computer Vision Tasks}
For \ac{cv} classification, we benchmark ResNet-50 and ViT-L/16 as representative convolutional and transformer-based architectures.
For \ac{cv} generation, we use Stable Diffusion v2 to capture the mixed compute and memory behavior of modern diffusion-model training.
To avoid data-loading and I/O bottlenecks, randomized RGB inputs are pre-generated and preloaded into GPU memory.
Classification uses 224\,$\times$\,224 inputs, whereas image generation uses 512\,$\times$\,512 inputs.

\paragraph{Large Language Model Tasks}
For \ac{llm} workloads, we benchmark LLaMA~3 8B in two modes, namely pre-training using LitGPT and inference serving using SGLang.
Pre-training is performed on an English text corpus comprising approximately 30,000,000 tokens.
The LLaMA~3 8B model weights are obtained from Hugging Face\footnote{\url{https://huggingface.co/meta-llama/Meta-Llama-3-8B}} and must be downloaded prior to benchmark execution.

\subsection{Measurement Protocol}

All benchmarks follow a standardized GPU-centric measurement protocol designed to eliminate non-compute bottlenecks and to support reproducible comparisons across systems. 
Each experiment is executed for 11 training epochs, with the first epoch discarded as a warm-up phase to remove initialization effects and the remaining 10 epochs used to average out short-term runtime fluctuations in throughput and energy measurements. 
Performance and power measurements are collected exclusively during active training batches via GPU-level callbacks, excluding any overhead.
Measurements of throughput and energy metrics are sampled at a fixed high-frequency interval of 100 ms.

The \ac{llm} workloads follow the same throughput-oriented principle, using tokens per second as the primary metric. 
For pre-training, warm-up steps are excluded from performance reporting to eliminate initialization transients and JIT compilation effects, and throughput and loss metrics are collected at each training step through framework-level logging callbacks. 
For inference, the benchmark server is fully warmed up prior to measurement.

\subsection{Software Environment}

Full container definitions and configuration files are available in the publicly accessible repository accompanying this work.

All computer vision experiments are deployed as Apptainer containers to ensure portability and reproducibility across \ac{hpc} systems.
In both cases, the most recently available container versions at the time of measurement were used.
For NVIDIA hardware, the container is built on top of the NVIDIA NGC base image.\footnote{\texttt{nvcr.io/nvidia/pytorch:25.05-py3}; PyTorch~2.8.0, Python~3.12, CUDA~12.9.0; additionally installed: xFormers~0.0.30 with support for FlashAttention kernels, PyTorch Lightning, \texttt{open-clip-torch}~2.7.0, \texttt{einops}, \texttt{omegaconf}, \texttt{pynvml}, and \texttt{scipy}.}
For AMD hardware, the container is based on the ROCm PyTorch image.\footnote{\texttt{rocm/pytorch:rocm7.1.1\_ubuntu24.04\_py3.12\_pytorch\_release\_2.9.1}; PyTorch~2.9.1, Python~3.12, ROCm~7.1.1; additionally installed: xFormers and \texttt{torchvision} from the PyTorch ROCm~7.1 wheel index, PyTorch Lightning, \texttt{open-clip-torch}~2.7.0, \texttt{einops}, \texttt{omegaconf}, and \texttt{scipy}; RDMA/\texttt{libibverbs} and OpenMPI~5.0.5 configured for InfiniBand and multi-node communication.}

All \ac{llm} experiments are deployed as Apptainer containers respectively.
For the pre-training benchmark, the container includes PyTorch with LitGPT and all required dependencies for distributed training. For NVIDIA hardware, the container is built on top of the NVIDIA NGC PyTorch base image with CUDA~12.4+ support. 
For AMD hardware, the container is based on the ROCm PyTorch image with ROCm~6.3.0+ and appropriate RDMA/InfiniBand configurations for multi-node communication.
For the inference benchmark, precompiled Docker/Apptainer containers are provided for both NVIDIA (CUDA~12.4+) and AMD (ROCm~6.3.0+) platforms, incorporating SGLang and all serving infrastructure. The containers include hardware-specific optimizations for mixed-precision inference and inter-node communication where applicable.

\subsection{Experimental Setup}

To study the power and performance trade-off of the considered \ac{ai} benchmarks, we execute each workload on nodes equipped with different GPU accelerators (\cref{tab:hardware}) and vary the accelerator power limit systematically. The resulting throughput and energy measurements are aggregated at the node level for each selected power-limit setting. We only report the power draw of the GPU and do not account for other energy consumers in the node.

We evaluated the two NVIDIA cards running in power state ``P0'', corresponding to the highest performance state, and the AMD MI300X in mode ``auto'', because the mode ``high'' was not selectable at the time of testing.
We use the highest batch size that remains stable for each workload and platform.

\begin{table*}[t]
    \centering
    \small
    \caption{Specifications of the tested GPU nodes.}
    \setlength{\tabcolsep}{5pt}
    \begin{tabular}{l | c c c}
        \toprule
         & \textbf{NVIDIA H100} & \textbf{NVIDIA H200} & \textbf{AMD MI300X} \\
        \midrule
        Memory capacity & 94 GB & 141 GB & 192 GB \\
        Memory bandwidth & 2.4 TB/s & 4.8 TB/s & 5.3 TB/s \\
        Memory interface & HBM2e & HBM3 & HBM3 \\
        TDP & \multicolumn{2}{c}{700 W} & 750 W \\
        Peak performance FP32 & \multicolumn{2}{c}{67 TFLOP/s} & 163.4 TFLOP/s \\
        Peak performance TF32 & \multicolumn{2}{c}{989 TFLOP/s} & 653.7 TFLOP/s \\
        GPUs per node & \multicolumn{2}{c}{4} & 8 \\
        GPU interconnect & \multicolumn{2}{c}{900 GB/s} & 896 GB/s \\
        Server & \multicolumn{2}{c}{Lenovo SD665-N V3} & Lenovo SR685a V3 \\
        Power state (set) & \multicolumn{2}{c}{P0} & auto \\
        \bottomrule
    \end{tabular}
    \label{tab:hardware}
\end{table*}

All runs were performed in a multi-GPU setting on a single node. For better comparability with other work, we report the values on a per-GPU basis. For the H100 and H200 platforms, one node comprises 4 GPUs, and all reported measurements therefore use 4 GPUs. The MI300X platform provides 8 GPUs per node. For a fair cross-platform comparison, the reported MI300X results are limited to a 4-GPU configuration. On AMD, power limits below 400~W were not reliably enforced and the GPUs drew more power than configured. The resulting software-stack stability issues are discussed in \cref{sec:software_stack}.

We sweep the GPU power limit from 200~W to 700~W in steps of 100~W. For completeness, we additionally report results at 750~W only for the AMD MI300X, which corresponds to its maximum power limit. 

\section{Performance Analysis}
\label{sec:results}

\subsection{Computer Vision Benchmark Results}
\label{subsec:results_cv}

\begin{table*}[t]
    \centering
    \caption{Samples per second per GPU for the computer vision workloads under different power limits.}
    \setlength{\tabcolsep}{.4em}
    \begin{tabular}{llrrrrrrr}
        \toprule
        Model & GPU & 200 W & 300 W & 400 W & 500 W & 600 W & 700 W & 750 W\\
        \midrule
        \multirow{3}{*}{ResNet-50}
            & H100 & \multicolumn{1}{r}{\underline{842.78}} & \multicolumn{1}{r}{\underline{1312.8}} & 1426.0 & 1463.2 & 1475.0 & 1479.3 & n/a \\
            & H200 & 650.93 & 1115.7 & \multicolumn{1}{r}{\underline{1619.2}} & \multicolumn{1}{r}{\underline{1915.1}} & \multicolumn{1}{r}{\underline{2061.4}} & \multicolumn{1}{r}{\underline{2148.6}} & n/a \\
            & MI300X & (771.68) & (789.94) & 924.17 & 1117.3 & 1312.8 & 1464.9 & \multicolumn{1}{r}{\underline{1525.1}} \\
        \midrule
        \multirow{3}{*}{ViT-L/16}
            & H100 & \multicolumn{1}{r}{\underline{116.40}} & \multicolumn{1}{r}{\underline{214.90}} & \multicolumn{1}{r}{\underline{259.00}} & 280.05 & 297.79 & 305.97 & n/a \\
            & H200 & 87.23 & 170.57 & 257.12 & \multicolumn{1}{r}{\underline{301.12}} & \multicolumn{1}{r}{\underline{331.10}} & \multicolumn{1}{r}{\underline{358.70}} & n/a \\
            & MI300X & (76.64) & (89.53) & 126.51 & 159.30 & 177.01 & 198.05 & \multicolumn{1}{r}{\underline{206.61}} \\
        \midrule
        \multirow{3}{*}{\begin{tabular}{@{}l@{}}Stable \\ Diff. v2\end{tabular}}
            & H100 & 20.14 & \multicolumn{1}{r}{\underline{34.25}} & 39.38 & 41.62 & 42.77 & 43.28 & n/a \\
            & H200 & 15.50 & 29.26 & \multicolumn{1}{r}{\underline{42.25}} & \multicolumn{1}{r}{\underline{49.37}} & \multicolumn{1}{r}{\underline{53.94}} & \multicolumn{1}{r}{\underline{57.30}} & n/a \\
            & MI300X & \multicolumn{1}{r}{(\underline{23.94})} & (25.13) & 30.19 & 37.48 & 43.23 & 47.86 & \multicolumn{1}{r}{\underline{48.87}} \\
        \bottomrule
    \end{tabular}
    \label{tab:results_cv_combined}
\end{table*}

\Cref{tab:results_cv_combined} reports per-GPU throughput across all three computer vision workloads for the evaluated power-limit settings.
As noted in the experimental setup, power limits below 400\,W were not enforced reliably on the AMD MI300X; we therefore interpret the MI300X results, particularly at low power limits, with caution.

For \ac{resnet}-50, the H100 improves rapidly from low to mid power limits and then shows diminishing returns at higher power. 
The H200 scales more steadily across the evaluated range and achieves the highest throughput at the upper power settings (2149 vs.\ 1479\,imgs/s for H100 at 700\,W). At high power limits, the H100 may become bandwidth-limited earlier due to its lower HBM bandwidth.
The MI300X increases more gradually and only becomes more competitive at the highest power limit, exceeding the H100 only at 750\,W.

For \ac{vit}-L/16, the qualitative trends are similar, but the MI300X lags more strongly, which is consistent with the larger sensitivity of attention-heavy transformer workloads to framework and kernel maturity.
At 700\,W, the H200 reaches 359\,imgs/s, whereas the MI300X reaches 198\,imgs/s.

\ac{ldm} exhibits a different ordering across power limits.
At low power, the MI300X can lead (23.9\,imgs/s at 200\,W), which is consistent with its large HBM capacity and bandwidth.
At higher power limits, the H200 achieves the highest throughput (57.3\,imgs/s at 700\,W) and, compared to the H100, likely benefits from its substantially higher HBM bandwidth.
This also suggests that, for the more memory-sensitive parts of these workloads, the H100 can become bandwidth-limited and therefore saturate earlier than the H200.

\subsection{Large Language Models Benchmark Results}
\label{subsec:results_llm}

\begin{table*}[b]
    \centering
    \caption{Tokens per second per GPU for the large language model workloads under different power limits.}
    \footnotesize
    \setlength{\tabcolsep}{.3em}
    \begin{tabular}{llrrrrrrr}
        \toprule
        Mode & GPU & 200\,W & 300\,W & 400\,W & 500\,W & 600\,W & 700\,W & 750\,W \\
        \midrule
        \multirow{3}{*}{\begin{tabular}{@{}l@{}}Pre-\\training\end{tabular}}
            & H100 & \multicolumn{1}{r}{\underline{9699.3}} & \multicolumn{1}{r}{\underline{17982.3}} & \multicolumn{1}{r}{\underline{24853.4}} & 27705.3 & 29250.0 & 30271.0 & n/a \\
            & H200 & 8914.9 & 14244.7 & 22285.7 & \multicolumn{1}{r}{\underline{28338.8}} & \multicolumn{1}{r}{\underline{31586.8}} & \multicolumn{1}{r}{\underline{34107.2}} & n/a \\
            & MI300X & (9200.7) & (11138.8) & 13119.8 & 14858.6 & 16391.0 & 16724.5 & \multicolumn{1}{r}{\underline{16843.3}} \\
        \midrule
        \multirow{3}{*}{Inference}
            & H100 & 780.5 & \multicolumn{1}{r}{\underline{1472.7}} & \multicolumn{1}{r}{\underline{1833.1}} & 2027.9 & 2154.5 & 2191.4 & n/a \\
            & H200 & 571.8 & 1255.0 & 1804.2 & \multicolumn{1}{r}{\underline{2063.7}} & \multicolumn{1}{r}{\underline{2289.7}} & \multicolumn{1}{r}{\underline{2417.5}} & n/a \\
            & MI300X & \multicolumn{1}{r}{(\underline{960.6})} & (1145.4) & 1400.1 & 1616.6 & 1758.7 & 1824.3 & \multicolumn{1}{r}{\underline{1849.7}} \\
        \bottomrule
    \end{tabular}
    \label{tab:results_llm_combined}
\end{table*}

\Cref{tab:results_llm_combined} reports per-GPU throughput in tokens per second for the pre-training and inference workloads.
For pre-training, the three accelerators exhibit clearly different scaling behavior across the tested power range. 
At low power, the H100 delivers the highest throughput and improves most sharply as the power limit increases. 
The H200 starts slightly behind, but scales more strongly toward higher power and overtakes the H100 in the middle of the range. 
At 700\,W, it achieves the best result overall with 34\,107\,tokens/s, compared with 30\,271\,tokens/s for the H100. 
By contrast, the MI300X improves more gradually across the full range and remains well below both NVIDIA accelerators even at its highest tested setting. 
Overall, the H100 appears strongest under tighter power budgets, whereas the H200 is the most effective choice when higher power limits are available.

Inference shows the same qualitative ranking as pre-training.
The H100 again performs best at low power and shows a strong gain as the power budget increases. 
The H200 begins below the H100, but scales better toward higher power and becomes the fastest NVIDIA accelerator in the upper part of the range, reaching 2\,418\,tokens/s at 700\,W versus 2\,191\,tokens/s for the H100. 
The MI300X is notable for its strong low-power inference performance, where it exceeds both NVIDIA GPUs, but this advantage disappears as the power limit increases because its scaling is comparatively limited. 
As a result, it falls behind both the H100 and H200 at higher power limits.

\section{Energy Efficiency Analysis}
\label{sec:discussion}

\begin{figure}[t]
    \centering
    \includegraphics*[width=\linewidth]{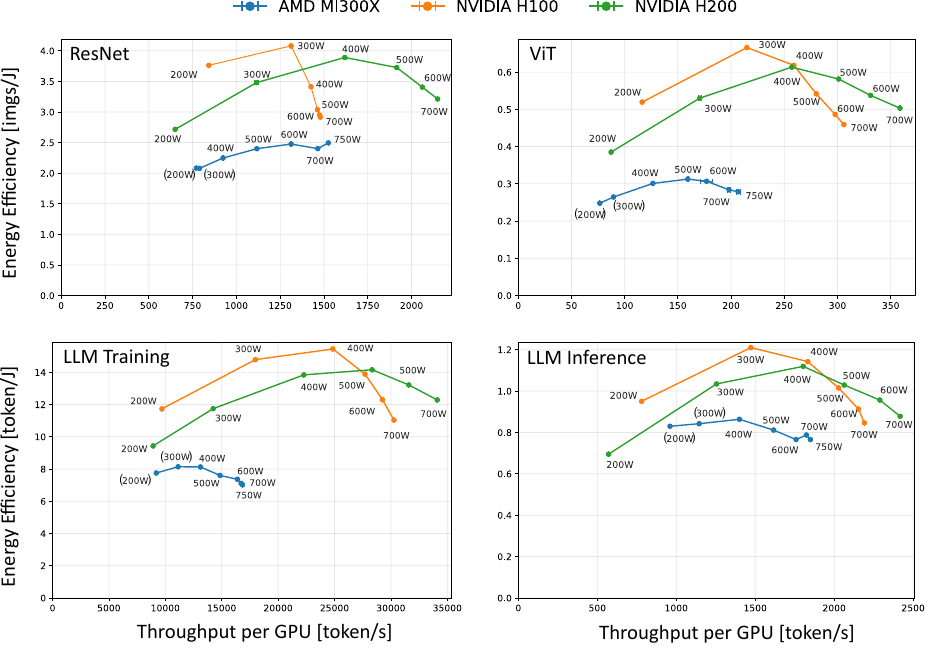}\vspace{-4mm}
    \caption{Energy efficiency of the \ac{ai} benchmarks at different power limits.}
    \label{fig:efficiency}
\end{figure}

\Cref{fig:efficiency} shows energy efficiency in ``work units per Joule'' versus performance, with the power limit as a parameter along each curve. This allows us to track performance and energy at the same time and clearly exposes the performance-energy Pareto front. 
We define energy efficiency as
\begin{equation}
\eta = \frac{T}{P_{\mathrm{GPU}} + \frac{100\,\mathrm{W}}{N_{\mathrm{GPU}}}} \, ,
\end{equation}
where $T$ denotes the application throughput in work units per second, $P_{\mathrm{GPU}}$ is the measured GPU power, and $N_{\mathrm{GPU}}$ is the number of active GPUs per node. The energy efficiency values assume an additional constant 100\,W per node as an estimated fixed system-level power overhead beyond the accelerators themselves.
Since all devices have a nonnegligible baseline power, energy efficiency must drop for very low power limits. At high power limits, the power-vs-frequency behavior is generally nonlinear (i.e., power grows faster than linear with frequency), which also leads to a drop in efficiency. This explains the hill-shaped trends visible in the data. 
The location and sharpness of the efficiency peak depend on both the workload and the GPU architecture, which indicates that no single power setting is optimal across all cases. 

For the H100, the efficiency peak is at 300\,W for \ac{resnet}-50, Stable Diffusion v2, and \ac{llm} inference, whereas \ac{vit}-L/16 and \ac{llm} pre-training peak at 400\,W. 
In several cases, especially for transformer-based workloads, increasing the limit from 300\,W to 400\,W yields additional throughput with only a modest reduction in efficiency. 
By contrast, the efficiency of \ac{resnet}-50 and Stable Diffusion v2 decreases more noticeably beyond 300\,W. 
Across the evaluated workloads, the H100 also attains the highest peak efficiency values among the three accelerators.

The H200 exhibits slightly lower peak efficiency than the H100, but its curves are flatter and show less pronounced efficiency loss at higher power limits. 
For most workloads, the efficiency optimum is located at 400\,W, while \ac{llm} pre-training peaks at 500\,W. 
This behavior suggests that operating the H200 somewhat above its efficiency optimum can still be reasonable when higher throughput is desired.

The MI300X shows the lowest overall energy efficiency in our measurements, but also the flattest response to increasing power limits. 
In other words, its efficiency deteriorates less sharply as the power limit rises. As shown in \cref{fig:tdp}, this behavior coincides with a power-limit-dependent change in memory clock on the AMD platform, which likely contributes to the observed trend.

\section{Discussion}
\label{sec:software_stack}
Moving beyond synthetic microbenchmarks, our \ac{ai} training evaluation highlighted a noticeable divergence between hardware specifications and realized throughput. Although the AMD MI300X theoretically matches or outperforms NVIDIA's H100 and H200 (Table \ref{tab:hardware}), it generally yields lower performance in our specific scenarios. This gap appears to be closely tied to the current maturity of the software stack. We observed notable initialization overheads, with the AMD framework frequently requiring up to 15 minutes just to start training. The tools also presented challenges: \texttt{amd-smi} did not allow power state modifications, and the MI300X did not respect power limits set below 400 W. Additionally, approaching theoretical peak throughput typically requires specialized numerical formats like TensorFloat-32 (TF32). Because frameworks such as PyTorch Lightning often default on AMD to more conservative precisions, practitioners must manually balance acceleration against potential impacts on numerical stability -- a trade-off we did not explicitly isolate due to time constraints.
Our results show that while AMD’s hardware is highly competitive on paper, it delivers lower performance than its competitor in real-world applications. In its present state, the AMD software ecosystem appears to be less mature than the NVIDIA environment.
Importantly, for high-cost accelerators, throughput losses may outweigh moderate efficiency gains, so power limits should balance energy efficiency, throughput, and acquisition cost.
\section{Conclusion}
\begin{figure}[tb]
    \centering
    \begin{subfigure}{0.5\textwidth}
        \centering
        \includegraphics[width=\linewidth]{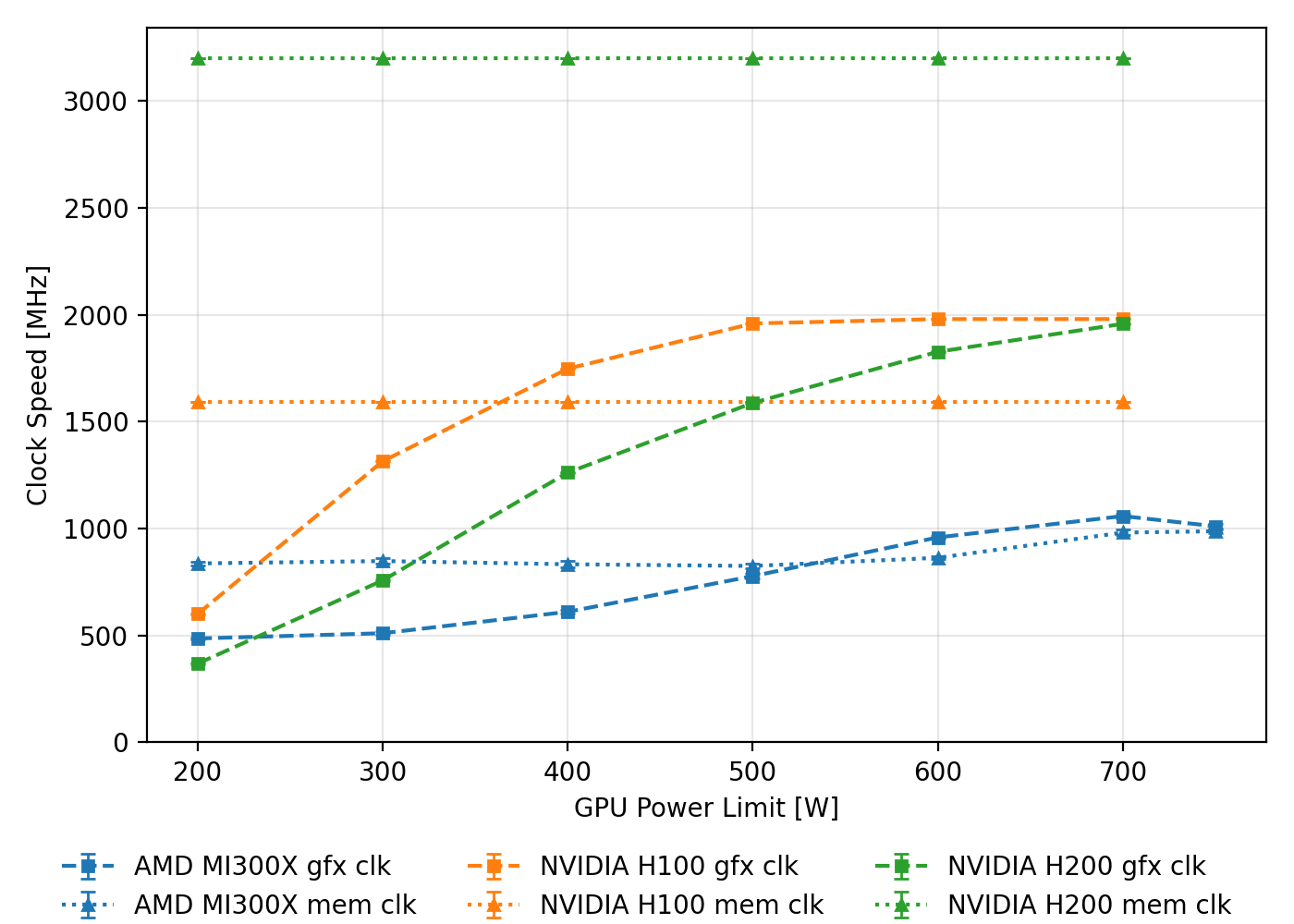}
        \caption{\ac{resnet}}
        \label{fig:sub1}
    \end{subfigure}\hfill
    \begin{subfigure}{0.5\textwidth}
        \centering
        \includegraphics[width=\linewidth]{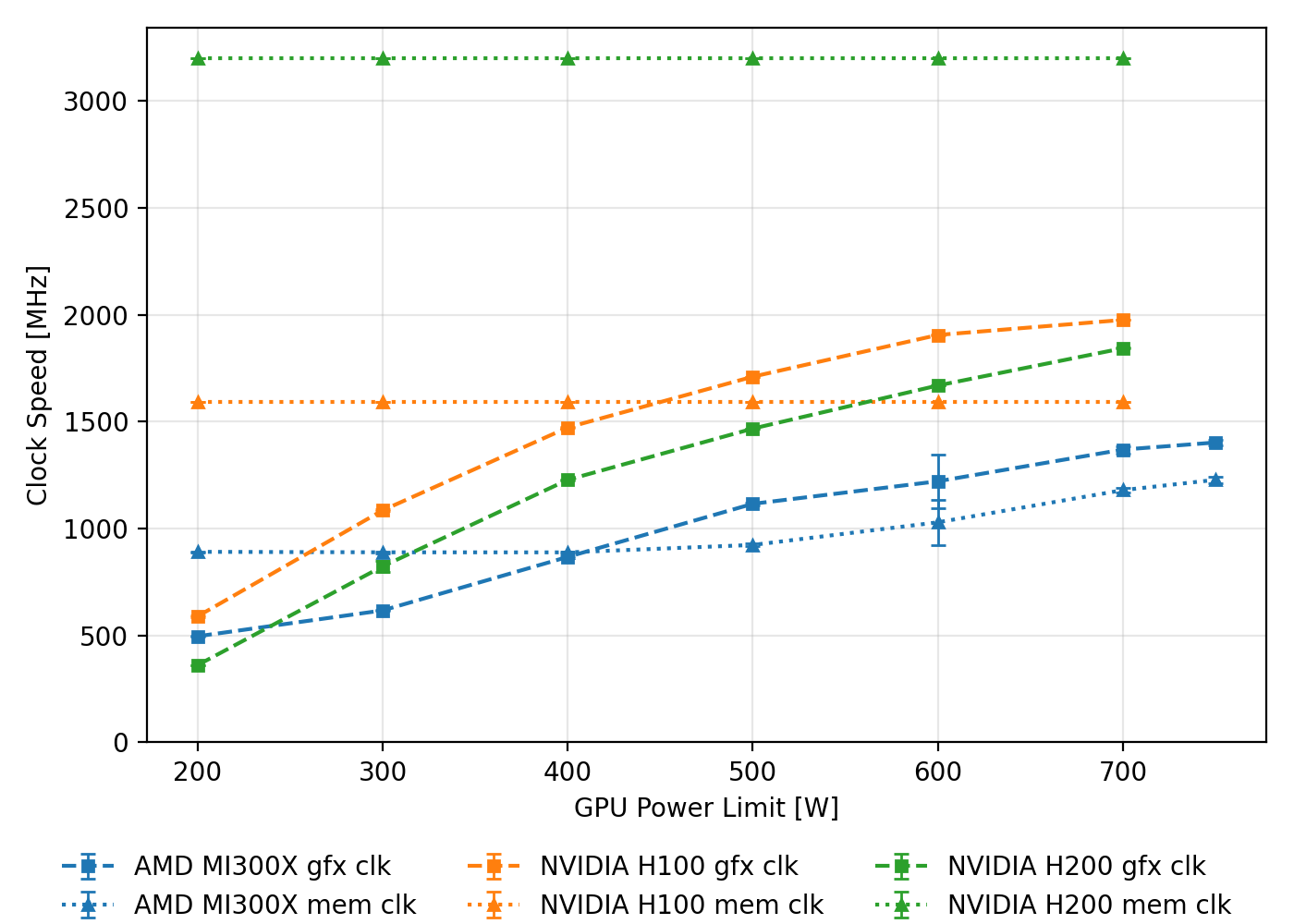}
        \caption{\ac{vit}}
        \label{fig:sub2}
    \end{subfigure}\hfill\vspace{-2mm}
    \caption{Clock frequency vs. power limit across two training workloads, i.e. \ac{resnet} and \ac{vit} for H100, H200, and MI300X.}
    \label{fig:tdp}
\end{figure}
 
Our analysis led to three major insights.
First, power limit and efficiency exhibit a characteristic hill-shaped relationship across all tested hardware, with suboptimal operating points at both very low and very high power limits.
Second, the H100 typically reaches its efficiency optimum around 300--400\,W depending on the workload, whereas the H200 reaches its optimum at slightly higher power budgets and shows flatter efficiency curves at high power.
We evaluate power limits relative to the thermal design power (TDP) of each GPU.
The clock-frequency trends in \cref{fig:tdp} are consistent with this picture: The H100 approaches saturation toward the upper end of the evaluated range, while the H200 does not yet show an equally clear plateau, suggesting that a higher TDP could still be beneficial, albeit likely at the cost of lower energy efficiency and higher cooling requirements.
Third, the AMD MI300X achieves low application performance (in relation to its theoretical hardware capabilities) in several workloads, most notably in transformer-based training and inference.
This indicates that the current ROCm/PyTorch software stack does not deliver the same level of hardware efficiency as the CUDA-based NVIDIA counterpart. Lacking AMD-specific optimizations in components such as xFormers may be a performance limiter. Also, the data presented for the low-power scenarios on AMD should be treated with caution because low AMD power limits could not be enforced reliably.
For \ac{hpc} center operators, these results imply that procurement and power-management decisions should be based on workload-specific throughput, energy efficiency, and software-stack maturity rather than on peak specifications alone.
\section{Outlook}
\label{sec:outlook_summary}

Several directions remain for future work.
Extending the benchmark to multi-node distributed training scenarios is a natural next step to assess inter-node communication overhead and efficiency under power constraints.
Another important direction is to align the \ac{cv} and \ac{llm} benchmark protocols more closely, for example with respect to warm-up handling and related measurement settings.
The impact of numerical precision, particularly the BF16 and the emerging FP8 formats, on throughput, energy efficiency, and hardware utilization also warrants dedicated investigation, because datatype selection is a first-order determinant of effective accelerator utilization.
The observed performance gap on the AMD platform further motivates a joint hardware and software analysis to identify specific bottlenecks in the ROCm/PyTorch stack and evaluate optimizations such as explicit BF16 casting, kernel selection, and profile-guided optimizations.
Finally, incorporating newer GPU generations such as NVIDIA Blackwell would enable longitudinal tracking of hardware efficiency trends in the \ac{hpc} \ac{ai} ecosystem.

\section*{Acknowledgment}

The authors gratefully acknowledge the scientific support and HPC resources provided by the Erlangen National High Performance Computing Center (NHR@FAU) of the Friedrich-Alexander-Universität Erlangen-Nürnberg (FAU) under the BayernKI project v111dc. BayernKI funding is provided by Bavarian state authorities.
The authors gratefully acknowledge the scientific support and HPC resources provided by NHR@FAU of the FAU under the NHR project b104dc. NHR funding is provided by federal and Bavarian state authorities. NHR@FAU hardware is partially funded by the German Research Foundation (DFG) – 440719683.

\bibliographystyle{splncs04}
\bibliography{lit_no_links}

\end{document}